\documentclass[twocolumn,aps,superscriptaddress,floatfix]{revtex4}
\usepackage{graphicx,epsfig,subfig,dcolumn,bm,mathrsfs,amsmath,color}
\begin{document}

\title{Dynamic and Dielectric Response of Charged Colloids \\
  in Electrolyte Solutions to External Electric Fields}

\author{Jiajia Zhou}
\email[]{zhou@uni-mainz.de} 
\affiliation{Institut f\"ur Physik, Johannes Gutenberg-Universit\"at Mainz, Staudingerweg 7, D-55099 Mainz, Germany}
\author{Roman Schmitz}
\affiliation{Max-Planck-Institut f\"ur Polymerforschung, Ackermannweg 10, D-55128 Mainz, Germany}
\author{Burkhard D\"unweg}
\affiliation{Max-Planck-Institut f\"ur Polymerforschung, Ackermannweg 10, D-55128 Mainz, Germany}
\affiliation{Department of Chemical Engineering, Monash University, Melbourne, VIC 3800, Australia}
\author{Friederike Schmid}
\email[]{friederike.schmid@uni-mainz.de} 
\affiliation{Institut f\"ur Physik, Johannes Gutenberg-Universit\"at Mainz,
 Staudingerweg 7, D-55099 Mainz, Germany}

 \begin{abstract}
Computer simulations are used to investigate the response of a charged colloid and its surrounding microion cloud to an external electric field. 
Both static fields (DC) and alternating fields (AC) are considered. 
A mesoscopic simulation method is implemented to account in full for hydrodynamic and electrostatic interactions.
The response of the system can be characterized by two quantities: the mobility and the polarizability. 
Due to the interplay of the electrostatic attraction and hydrodynamic drag, the response of the microions close to the colloid surface is different from that of the microions far away from the colloid. 
Both the mobility and polarizability exhibit a dependency on the frequency of the external fields, which can be attributed to the concentration polarization, the mobility of the microions, and the inertia of microions. 
The effects of the colloidal charge, the salt concentration, and the frequency of the external fields are investigated systematically.
 \end{abstract}


\maketitle

\section{Introduction}
\label{sec:introduction}

Colloidal suspensions have numerous applications in different fields
such as chemistry, biology, medicine, and engineering. In an aqueous
solution, colloidal particles are often charged, either by
ionization or dissociation of a surface group, or preferential
adsorption of ions from the solution. Therefore, they can be
efficiently controlled or manipulated by electric fields. Electric
fields can be applied in two different ways. One is to use static
(DC) fields which induce electrophoresis, \emph{i.e.}, the migration of
individual colloids \cite{RSS, Hiemenz_colloid3}. Electrophoresis is
commonly used to measure the surface charge density of colloidal
particles. The other possibility is alternating electric field (AC
field). The time-dependent perturbation allows one to selectively
probe dynamic phenomena at different time scales and produces
substantially  more information than just probing the response to a
static field \cite{RSS, Dhont}.

The colloids respond to external fields on relatively short time
scales and in an often fully reversible way. One important quantity
characterizing the dynamic response of the colloid is the
electrophoretic mobility. If an external electric field of the form
$\mathbf{E} = E_0 \exp(i \omega t) \hat{\mathbf{x}}$ is applied in
the $x$-direction, the $x$-component of the colloid velocity
exhibits an oscillation of the form $v_x = v_0 \exp(i \omega t)$,
where $v_0$ can be a complex number. For weak fields, the frequency
of the colloid motion is the same as the external field, and the
response depends linearly on the strength of the perturbation. The
amplitude of the colloid velocity then can be written as
 \begin{equation}
  \label{eq:mu_def}
  v_0 = \mu(\omega) E_0.
 \end{equation}
The mobility $\mu(\omega)$ is in general a complex number, whose
magnitude characterizes the sensitivity of the response,
\emph{i.e.}, how fast the colloid moves under fixed external field,
and the phase describes the synchronization between the response and
perturbation, \emph{i.e.}, whether the colloid motion is in phase
with the external field. We explicitly include the argument $\omega$
to emphasize that the mobility is a function of the driving
frequency.

The microions, which include the counterions from the surface charge
and the dissolved salt molecules in the solution, play an important
role in determining the colloidal response to the external field.
Counterions accumulate around the colloid surface due to the
electrostatic attraction between opposite charges and form an
electric double layer (EDL). In close vicinity of the colloid, the
counterions stick to the surface and move together with the colloid.
Further away, microions are relatively mobile and experience thermal
motion. The thickness of the electric double layer is characterized
by the Debye screening length
 \begin{equation}
  \label{eq:Debye}
  l_D = \kappa^{-1} = \left[ 4 \pi l_B \sum_i z_i^2 \rho_i(\infty)
  \right]^{-\frac{1}{2}},
 \end{equation}
where the summation runs over different ion types. In Eq.~(\ref{eq:Debye}), $l_B=e^2/(4\pi\epsilon_m k_BT)$ is the Bjerrum
length which depends on the medium permittivity $\epsilon_m$ and the
temperature of the solution $T$, and $z_i$ and $\rho_i(\infty)$ are
the valence and bulk concentration for type $i$ ion, respectively.
When an external electric field is applied to the suspension, both
the colloidal particle and its surrounding electric double layer
will be polarized. The colloid acquires a dipole moment of the form
$p_0 \exp(i \omega t)$ in the direction of the applied field, and
the amplitude the dipole moment can be written as
 \begin{equation}
  \label{eq:alpha_def}
  p_0 = \alpha(\omega) E_0 .
 \end{equation}
The polarizability $\alpha(\omega)$ is used to characterize the
colloid's dielectric response to the external fields, and it is a
complex function of the frequency and the amplitude of the external
field. In the linear response region, the dipole moment is
proportional to the magnitude of the external field; thus the
polarizability does not depend on the field strength for weak
external fields.

The response of the colloid to the external field, be it dynamic
(\ref{eq:mu_def}) or dielectric (\ref{eq:alpha_def}),
is a combined effect of hydrodynamic and electrostatic interactions.
We can control the response by adjusting various factors such as the
surface charge density of the colloid and the salt concentration.
Fischer \emph{et al.} observed a salt-dependent change of sign in
the mobility of the condensed counterions in polyelectrolyte
solutions \cite{Fischer2008}. Similar phenomena are observed in our
colloidal system, and the mobility change can also be induced by
varying the colloid charges. Furthermore, we explore the response to
alternating electric fields, and systematically study the
frequency-dependence of mobility and polarizability.

Various theories have been proposed to understand the
frequency-dependent response of colloid dispersions. Already for
uncharged colloids, this response is far from trivial
\cite{Dhont2010,2012_q0}. On the MHz scale, the main contribution
to the induced dipole moments stems from the conductivity mismatch
between the particle and the solvent due to the presence of the free
microions in the solvent. The so-called Maxwell-Wagner relaxation
\cite{Maxwell1954, Wagner1914} only depends on the bulk properties
of the solution and the colloid, and has been widely used to
interpret experimental results \cite{Green1999, Ermolina2005}. For
charged colloids, the electric double layer plays an important role,
and its contribution can be included in the Maxwell-Wagner theory as
a surface conductivity term, which was first introduced by O'Konski
\cite{OKonski1960}. The Maxwell-Wagner-O'Konski theory has also been
extended to ellipsoidal colloids \cite{Saville2000}. On long time
scales up to 100 MHz, a salt concentration gradient builds up along
the colloid and the thickness of the double layer varies
accordingly, leading to an additional source of polarization
($\alpha$-polarization). Theories for the low-frequency response
have been developed in the Ukraine school \cite{DukhinShilov} based
on the standard electrokinetic model. Their analytic results rely on
the assumption that the electric double layer is much thinner than
the radius of the colloid. This is justified for micrometer-sized
colloids, but becomes questionable for particles of nanometer
radius. For situations that involve thick electrical double layers
and the whole frequency spectrum, one can solve the full
electrokinetic equations using various numerical methods
\cite{OBrien1978,DeLacey1981, Hill2003, Kim2006, Nakayama2008,
Zhaohui2009,Schmitz2012}.

Using molecular dynamic simulations to study macroion solutions
under alternating electric fields is a relatively new approach; only
a few works in the literature have tackled this subject. Most
studies have focused on the conformation change of a single
polyelectrolyte (PE) chain. Liu \emph{et al.} studied the unfolding
and collapse of a flexible PE chain under a sinusoidal electric
field \cite{LiuHongjun2010}. Hsiao \emph{et al.} examined a similar
system, but in trivalent salt solutions under a square-wave electric
field \cite{HsiaoPai-Yi2011}. Zhang \emph{et al.} investigated the
dynamics of an anchored PE chain, for both flexible and semiflexible
cases \cite{ZhangQi-Yi2012}. The detection of DNA sequences using an
AC-field in a nanopore capacitor was discussed in Ref.
\cite{Sigalov2008}. A Langevin thermostat was used in most studies,
because the hydrodynamic interactions were taken to be screened for
long PE chains in solutions with high salt concentration. For the
system of nanometer-sized colloids, hydrodynamic interactions are
important, and we include coarse-grained solvent particles.

Simulations with explicit solvents and microions are numerically
challenging, because both the hydrodynamic and the electrostatic
interactions are long-ranged. In recent years, a number of
coarse-grained simulation methods have been developed to address
this class of problem. The general idea is to couple the explicit
charges with a mesoscopic model for Navier-Stokes fluids. There are
a few choices of the fluid model in the literature, such as the
Lattice Boltzmann (LB) method \cite{Lobaskin2004, Lobaskin2004a,
Lobaskin2007, Chatterji2005, Chatterji2007, Giupponi2011},
Multi-Particle Collision Dynamics (MPCD) \cite{Malevanets1999,
Gompper2009}, and Dissipative Particle Dynamics (DPD)
\cite{Hoogerbrugge1992, Espanol1995, Groot1997}. In this paper, we use
the particle-based DPD approach. DPD is a coarse-grained simulation
method which is Galilean invariant and conserves momentum. Since it
is a particle-based method, microions can be introduced in a
straightforward manner. A recent comparative study
\cite{Smiatek2009} indicated that the electrostatic interaction is
the most expensive part in terms of the computational cost.
Therefore, for intermediate or high salt concentrations, different
methods for modelling the fluid becomes comparable.

In this work, we study the response of the charged colloidal
particle and its surrounding ionic clouds under external electric
fields. In a recent publication \cite{2012_ac}, we have presented
first results on the frequency-dependent dielectric response of a
colloid in solutions of high ionic strength. In that case, the
results were in qualitative agreement with the prediction of the
Maxwell-Wagner theory. Here, we systematically vary the salt
concentration down to the low-salt regime where the Maxwell-Wagner
theory no longer applies, and analyze the contributions of microions 
to the dynamic and dielectric response. We use DPD
simulations, including in full the hydrodynamic and electrostatic
interactions. The remainder of this article is organized as follows:
In section \ref{sec:model}, we give a brief introduction to the
simulation model and describe important parameters for the system.
We present the simulation results on the mobility and polarizability
in section \ref{sec:mu} and \ref{sec:alpha}, respectively. Finally,
we conclude in section \ref{sec:summary} with a brief summary.

\section{Simulation Model}
\label{sec:model}

In this section, we briefly review our simulation model for a
colloidal particle in a salt solution and describe some important
physical quantities (see also Ref.\ \cite{2012_ac}). In the
following, physical quantities will be reported in a model unit
system of $\sigma$ (length), $m$ (mass), $\varepsilon$ (energy),
$e$(charge) and a derived time unit
$\tau=\sigma\sqrt{m/\varepsilon}$.

\begin{table*}[htbp]
\centering
\begin{tabular}{l p{0.4cm} l c}
\hline
fluid density $\rho$ & & 3.0 & $\sigma^{-3}$ \\
friction coefficient for fluid DPD interaction $\gamma_{DPD}$ & &
5.0 & $m/\tau$ \\
cutoff for fluid DPD interaction & & 1.0 & $\sigma$\\
shear viscosity $\eta_s$ & & $1.23 \pm 0.01$ & $m/(\sigma\tau)$ \\
friction coefficient for colloid DPD interaction & & 10.0 & $m/\tau$ \\
cutoff for colloid DPD interaction & & 1.0 & $\sigma$ \\
size of fluid particle, microion, and colloid & & 1.0, 1.0, 3.0 & $\sigma$ \\
mass of fluid particle, microion, and colloid & & 1.0, 1.0, 100 & $m$ \\
Bjerrum length $l_B$ & & 1.0 & $\sigma$ \\
scaled temperature $k_B T$ & & 1.0 &$\varepsilon$ \\
simulation time step $\Delta t$ & & 0.01 & $\tau$
\\
simulation box & & 30 & $\sigma$ \\
\hline
\end{tabular}
\caption{Parameters used in the DPD simulations.}
\label{tab:parameters}
\end{table*}

Our simulation system has three components: the solvents, the
microions, and the colloidal particle. The solvent is modelled as a
fluid of DPD beads, where DPD is used as a canonical thermostat
({\it i.e.} including the dissipative and stochastic part) without
conservative forces \cite{Soddemann2003}. Some important parameters
in the simulation are listed in Table~\ref{tab:parameters}. Our DPD
fluid has a density $\rho=3.0\,\sigma^{-3}$ and a shear viscosity
$\eta_s = 1.23 \pm 0.01\,m/(\sigma\tau)$. Counterions and salt
microions are introduced as the same DPD beads as the fluid
particle, but carry charges and have exclusive interactions. We only
consider the monovalent case where microions carry a single
elementary charge  $\pm e$. The exclusive interaction is necessary
to prevent the collapse of charged system. A short-range repulsive
Weeks-Chandler-Andersen interaction is used \cite{WCA},
 \begin{equation}
  \label{eq:WCA}
  V(r) = \begin{cases}
    4\varepsilon \left[ \left( \dfrac{\sigma}{r-r_0} \right)^{12} - \left(
      \dfrac{\sigma}{r-r_0} \right)^6 + \dfrac{1}{4} \right] & \text{for }
    r<r_c \\
    0 & \text{otherwise}
  \end{cases}
 \end{equation}
The cutoff radius is set at the potential minimum $r_c=r_0 +
\sqrt[6]{2}\,\sigma$. The microions have a size of $1.0\,\sigma$
($r_0=0$).

\begin{figure}[htbp]
  \includegraphics[width=0.9\columnwidth]{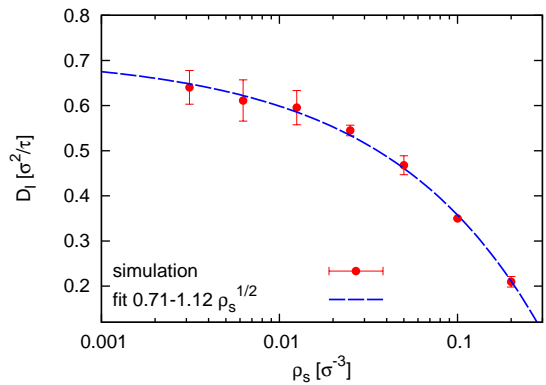}
  \caption{The diffusion constant of microions $D_I$, as a function of salt concentration $\rho_s$. The curve is a fit to Kohlrausch law with fitting parameters $A=0.71$ and $B=1.12$.}
  \label{fig:DI}
\end{figure}

One useful quantity for later discussions is the diffusion constant
of the microion $D_I$, which can be determined by measuring the
mean-square displacement
 \begin{equation}
  \lim_{ t \rightarrow \infty} \langle ({\bf r}(t)-{\bf r}(0))^2
  \rangle = 6 D_I t .
 \end{equation}
The diffusion constant is a function of the salt concentration. We
performed simulations with different salt concentrations,
$\rho_s=0.003125\,\sigma^{-3}$ -- $0.2\,\sigma^{-3}$. The simulation
results are compared with the empirical Kohlrausch law
\cite{Wright}, which states that microion diffusion constant depends
linearly on the square root of the salt concentration
$\sqrt{\rho_s}$,
 \begin{equation}
  D_I = A - B \sqrt{\rho_s},
\end{equation}
where $A$ and $B$ are fitting parameters.
Fig.~\ref{fig:DI} shows the simulation results and a fit to Kohlrausch law.

We model the large colloidal particle as a sphere with many
interacting sites on its surface. The interaction between these
surface sites and the solvent beads is modelled using DPD
dissipative interactions. To prevent solvent particles from
penetrating the colloid, a similar repulsive interaction as Eq.~(\ref{eq:WCA}) acts between the  colloid and fluid/microion
particles, but with the colloid radius $R=r_0+\sigma=3.0\,\sigma$.
The colloid carries a positive charge $Q$, and the mass of the
colloidal particle is $M=100\,m$ The moment of inertia is
$I=360\,m\sigma^2$, corresponding to a sphere with constant volume
density. The diffusion constant of an uncharged colloid is measured
by a linear regression of the mean-squared displacement, similar to
the case of microions. For a colloid with radius $R=3.0\,\sigma$,
the diffusion constant is measured to be $D=0.010 \pm
0.002\,\sigma^2/\tau$ in a simulation box with linear dimension
$L=30\,\sigma$. This result compares well to the diffusion constant
of a Stokes sphere of radius $3.0\,\sigma$ in a simple cubic lattice
\cite{Hasimoto1959}, $D=(k_BT/6\pi \eta_s)(1/R-2.837/L) =
0.010\,\sigma^2/\tau$.

We carried out simulations using the open source package ESPResSo
\cite{ESPResSo}. Modifications have been made to incorporate an
external time-dependent electric field. The temperature of the
system is kept at $k_B T = 1.0\,\varepsilon$. Electrostatic
interactions are calculated using the Particle-Particle-Particle
Mesh (P3M) method \cite{HockneyEastwood, Deserno1998, Deserno1998a},
with the Bjerrum length $l_B$ set to $1.0\,\sigma$. The
Velocity-Verlet algorithm is used to integrate the equation of
motion \cite{Verlet1967, Swope1982, FrenkelSmit} 
with a time step of $\Delta t = 0.01\,\tau$. 
We use a cubic simulation box with linear dimension
$L=30\,\sigma$, with periodic boundary conditions in all three
directions. We note that the application of periodic boundary
conditions makes the measurement in our system somewhat different
from the experimental situation. The periodic images of the colloid
are coupled both hydrodynamically and electrostatically. Therefore,
we measure in fact the mobility and polarizability of a particle in
a simple cubic lattice.

\begin{figure}[htbp]
  \includegraphics[width=1.0\columnwidth]{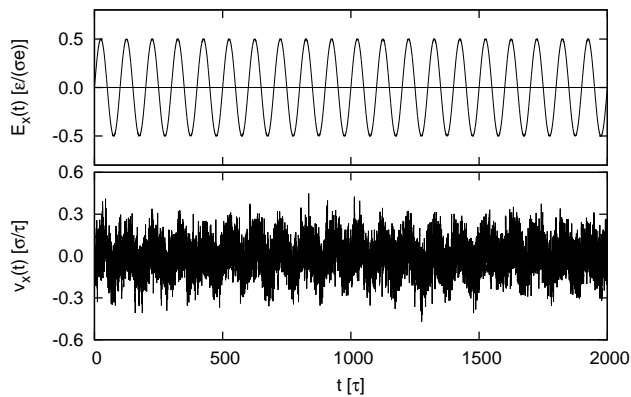}
  \caption{Dynamics of a charged colloidal particle with charge $Q=+50\,e$ and radius $R=3.0\,\sigma$ in a salt solution ($\rho_s=0.0125\,\sigma^{-3}$) under the influence of an AC field with frequency $f = 0.01\,\tau^{-1}$ and amplitude $E_0 = 0.5 \varepsilon / (\sigma e)$. Top: instantaneous electric field. Bottom: translational velocity along the field direction. }
  \label{fig:7t2_vel}
\end{figure}

In simulations, various time-series of the velocity or dipole moment
are obtained, but the data is noisy due to the thermal fluctuations.
For example, in Fig.~\ref{fig:7t2_vel} we show the external electric
field and the colloid velocity as a function of the time for one set
of parameters ($Q=50\,e$ and $\rho_s=0.0125\,\sigma^{-3}$). In order
to extract the amplitude and the phase of the velocity oscillation,
we apply a Fourier transform to the time-series of the velocity. The
real and imaginary parts of the Fourier spectrum exhibit peaks at
the frequency of the external electric field, and the peak values
correspond to the real and imaginary part of the complex velocity
amplitude.

To map the simulation units to real physical numbers, we use an
aqueous solution of KCl as a reference system. The energy unit
$\varepsilon$ is the thermal energy at room temperature
$k_BT=4.1\times 10^{-21}\,\mathrm{J}$. We set the Bjerrum length
$l_B=1.0\,\sigma$, thus the length unit $\sigma$ corresponds to the
Bjerrum length of water at room temperature,
$l_B=0.71\,\mathrm{nm}$. A salt concentration $0.05\,\sigma^{-3}$ in
simulation translates to an experimental value of
$0.228\,\mathrm{mol/L}$. We further equate the diffusion constant of
microions at zero concentration ($D^0_I=0.71 \,\sigma^2/\tau$) to
that of KCl in an aqueous solution \cite{Padding2006}. The diffusion
constant for K$^{+}$ and Cl$^-$ differ slightly ($D_{\rm
K^+}=1.96\times 10^{-9}\,\mathrm{m^2s^{-1}}$ and $D_{\rm
Cl^-}=2.03\times 10^{-9} \,\mathrm{m^2s^{-1}}$)
\cite{Hill2003,Ohshima}, so we use the value of $2.0\times 10^{-9}\,
\mathrm{m^2s^{-1}}$ for mapping. One simulation time unit then
corresponds to the real time $1.81\times 10^{-10}\,\mathrm{s}$.
Therefore, the frequency $f=0.1\,\tau^{-1}$ corresponds to an
experimental frequency 553 MHz.

\section{Dynamic response}
\label{sec:mu}

In the following sections, we report simulation results for a single
charged colloid of radius $R = 3.0\,\sigma$ in an electrolyte solution. 
We focus on the dynamic response of the colloid and microions,
characterized by the mobility $\mu$, and the dielectric response,
represented by the polarizability $\alpha$. We systematically vary
some important parameters, such as the salt concentration, the
colloid charge, and the frequency of the external fields.

\subsection{Effect of salt concentration}

To study the effect of the salt concentration and the colloid
charge, we will focus on the case of static electric fields. For
static fields, the motion of the colloid and its surrounding
microions can be characterized by a scalar mobility $\mu$, which
represents the ratio between the velocity of the particle and the
magnitude of the applied field. For the microions, the mobility also
depends on their position with respect to the colloid. In the
simulation, we average over the velocities of microions which have a
distance $r$ from the colloid center.

\begin{figure}[htbp]
  \includegraphics[width=1.0\columnwidth]{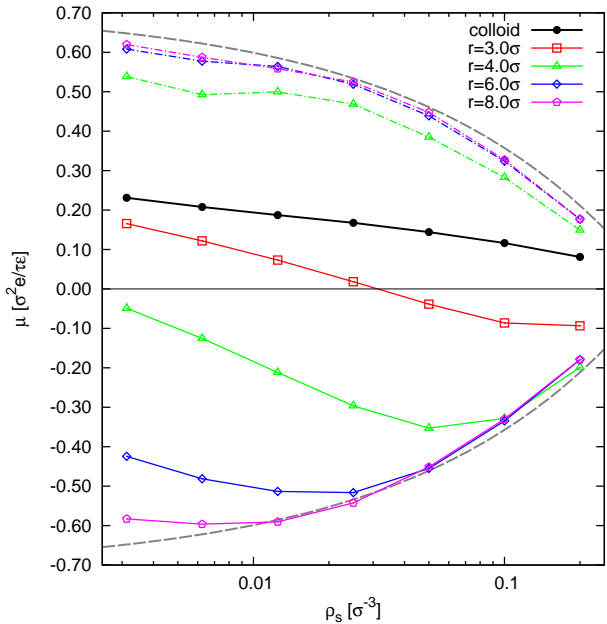}
  \caption{The mobility of the colloid and its surrounding microions with distance $r$ from the colloid center. The solid and dash-dotted lines correspond to the counterions and the coions, respectively. The mobilities are plotted as a function of the salt concentration $\rho_s$. The colloid has a bare charge of $Q=+50\,e$. Due to the electrostatic interactions, coions are excluded in the close vicinity of the colloid, therefore the mobility for coions at $r=3.0\,\sigma$ is absent. Also shown is the mobility of free microions in a salt solution without the colloid (dashed grey lines).}
  \label{fig:salt_dc}
\end{figure}

In Fig.~\ref{fig:salt_dc}, we plot the mobility of the colloid and
microions as a function of the salt concentration $\rho_s$. The
colloid carries a total charge of $Q=+50 e$. The number of
counterions to neutralize the colloid charge is comparable to the
number of salt at the lowest salt concentration, but is negligible
at intermediate and high salt concentrations. For the parameter
range considered here ($0.003125\,\sigma^{-3} \le \rho_s \le
0.2\,\sigma^{-3}$), the electrophoretic mobility of the colloid
depends only weakly on the salt concentration. When the salt
concentration increases, the Debye length becomes shorter, and the
electrostatic interaction becomes more screened. This reduces the
effective charge of the colloid, which includes the bare colloid
charge and the contribution from condensed counterions, resulting in
a slight reduction of the electrophoretic mobility of the colloid.

The mobilities for counterions at different distances $r$ away from
the colloid center are plotted in solid curves. Due to
the strong electrostatic attraction towards the colloid, counterions
overcome the entropic repulsion and accumulate in the vicinity of
the colloidal surface. Their mobility is a result of the interplay
between two conflicting interactions. On the one hand, counterions
have the opposite charge as the colloid, thus the external electric
field drives counterions away from the colloid. Similarly, the
entropic force also tends to separate counterions and the colloid
and pushes counterions towards regions where the ionic concentration
is lower. On the other hand, the Coulomb attraction from the colloid
charge and the hydrodynamic interaction from the colloid surface
drag counterions along with the colloid. The mobility of the
counterion is a result from these two competing effects. At low salt
concentration, counterions close to the colloid experience strong
attraction from the colloid, and they can be considered as attached
to the colloid surface. Therefore, the counterion mobility has the
same sign as that of the colloid. Note that both the hydrodynamic
and the Coulomb interaction depend on the distance between the
counterion and the colloid. When the distance is increased, both
interactions become weaker, resulting in a reduction of the
counterion mobility.

The counterion mobility also depends on the salt concentration. When
the salt concentration increases, the Debye length decreases and the
electrostatic interaction becomes screened, resulting in a reduction
of the attractive force from the colloid. In Fig.~\ref{fig:salt_dc}, all curves of counterion mobility exhibit an
initial reduction as the salt concentration increases. For
counterions close to the colloid surface, there exists a salt
concentration at which the electrostatic interaction is screened so
much that the external electric field dominates, resulting in a
change of the sign of the mobility. In this case, the external field
is strong enough to overcome the attraction, and drives even the
closest counterions to move in the opposite direction of the
colloid. A similar phenomenon has been discussed in Ref.
\cite{Fischer2008} for condensed counterions in polyelectrolyte
solutions.

The situation is different for coions, as shown in Fig.~\ref{fig:salt_dc} (dash-dotted lines). 
The coions have the same charge as the
colloid, thus the external fields and the hydrodynamic interactions
both drive the coions to move along with the colloid. The mobility
for the coions always has the same sign as the colloid, and its
dependence on the salt concentration shows the opposite trend,
compared to that of the counterions.

At high salt concentration and far away from the colloid, the
electrostatic and hydrodynamic interactions are fully screened, and
the forces exerted by the colloid on the microions are negligible.
In this case, the mobilities for counterions and coions are both
proportional to the diffusion constant of the microion $D_I$, but
with the opposite sign. 
The mobility for microions in solution without the colloid is shown as the dashed lines in Fig.~\ref{fig:salt_dc}.
The decrease of the mobility value at high salt
concentration is the result of the reduction in the diffusion
constant of the microion.

\subsection{Effect of colloid charge}

\begin{figure}[htbp]
  \includegraphics[width=1.0\columnwidth]{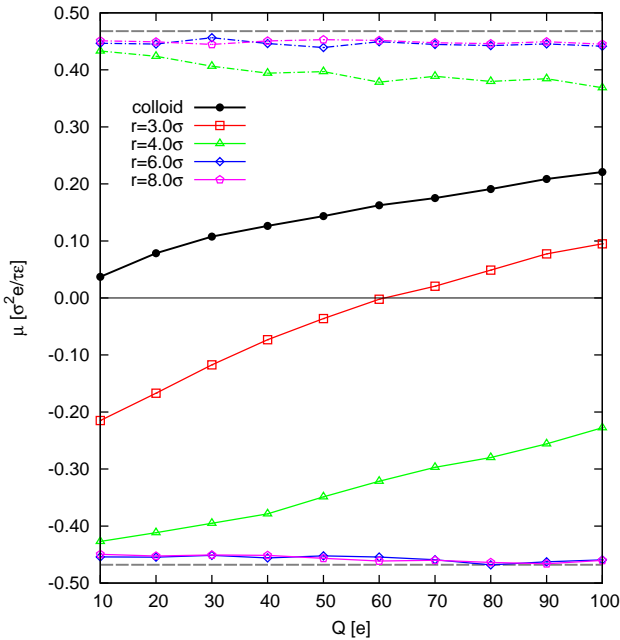}
  \caption{The mobility of the colloid and its surrounding microions with distance $r$ from the colloid center. The mobility of the counterions (solid lines) and coions (dash-dotted lines) are plotted as a function of the colloid charge. Due to the electrostatic interactions, coions are excluded in the close vicinity of the colloid surface, therefore the mobility for coions at $r=3.0\,\sigma$ is absent. The salt concentration is $\rho_s=0.05\,\sigma^{-3}$. Also shown is the microion mobility without the colloid (dashed grey lines), which is a constant for fixed salt concentration.}
  \label{fig:charge_dc}
\end{figure}

The response of the colloid and microions can also be tuned by
varying the bare charge on the colloid. Compared to the previous
situation of adjusting the salt concentration, where the range of
the electrostatic interaction (the Debye screen length) is varied,
the change of the colloid charge directly alters the strength of the
electrostatic interaction with the microions. Fig.~\ref{fig:charge_dc} shows the mobility of the colloid and its
surrounding microions as a function of the colloid charge $Q$. The
salt concentration is kept constant $\rho_s=0.05\,\sigma^{-3}$. The
electrophoretic mobility of the colloid increases as the colloid
charge increases, but the increase is less than linear.

The counterions are negatively charged, thus their mobility is in
general negative. The mobility depends strongly on the distance
between the counterions and the colloid. Far away from the colloid,
the electrostatic interaction is sufficiently screened for the salt
concentration $\rho_s=0.05\,\sigma^{-3}$. It can be seen in Fig.~\ref{fig:charge_dc} for the counterion curve $r=6.0\,\sigma$; the mobility
is insensitive to the colloid charge and almost coincides with the
mobility value for free negative ions (dashed grey line). Close to
the colloid surface, the attraction from the colloid charge and the
hydrodynamic drag dominate, and the difference between mobilities of
the counterion and the colloid becomes smaller. For highly charged
colloid $Q>60\,e$, the attraction is strong enough for the closest
counterions to be dragged along with the colloid, resulting in a
positive counterion mobility.

The mobility of coions (Fig.~\ref{fig:charge_dc}, dash-dotted lines) has the opposite sign from that of the counterions. 
Coions close to the colloid are slowed down by the slower colloid. 
The mobility of coions far from the colloid approaches that of free microions.

\subsection{Effect of frequency}

\begin{figure}[htbp]
  \includegraphics[width=1.0\columnwidth]{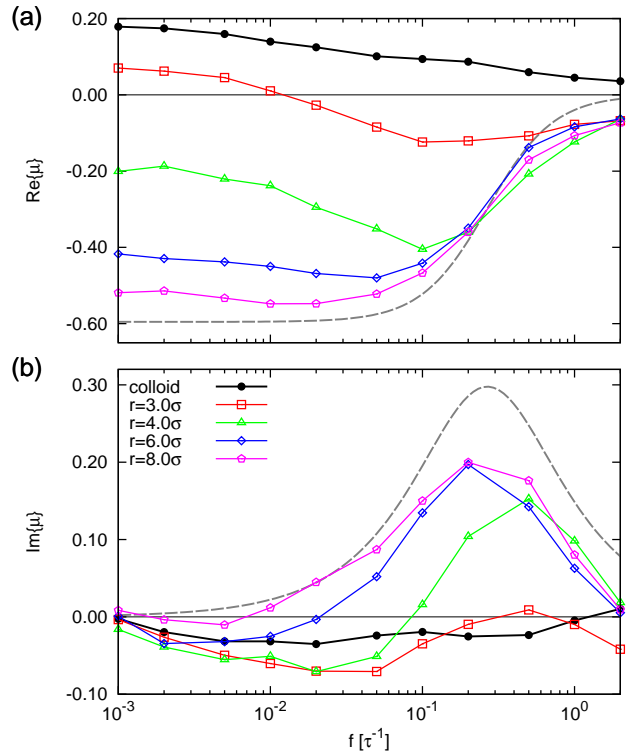}
  \caption{The mobility of the colloid and its surrounding counterions with distance $r$ from the colloid center. The real (a) and imaginary (b) components of the mobility are plotted as a function of electric field frequency $f=\omega/2\pi$. The colloid has a bare charge of $+50\,e$ and the salt concentration is $\rho_s=0.0125\,\sigma^{-3}$.}
  \label{fig:7t2_cn}
\end{figure}

In this section, we investigate the frequency-dependent response of the colloid in electrolyte solutions. 
We apply a sinusoidal external field to probe the colloidal suspension, and measure the mobility of the colloid and its surrounding counterions. 
The motion is in general not in phase with the external field. 
Thus, the mobility becomes a complex number, where the real part of the mobility characterizes the in-phase component of the motion, and the imaginary part is the out-phase contribution. 
In Fig.~\ref{fig:7t2_cn}, we show the complex mobility $\mu(\omega)$ for a system with salt concentration $\rho_s=0.0125\,\sigma^{-3}$ and colloid charge $Q=+50\,e$.

Let us focus on the counterions far away from the colloid first. 
For those counterions, the influence from the colloid is small; thus they can be approximated as free microions in a salt solution. 
The only important time scale is determined by the microion's inertia.
The equation of motion for a charged particle with mass $m$ and charge $-e$ in a viscous fluid is
 \begin{equation}
  m \ddot{x} = - \gamma \dot{x} - eE_0 e^{i\omega t},
 \end{equation}
where $\gamma=k_BT/D_I$ is the Langevin friction coefficient. 
The equation can be solved using Fourier transform, and one obtains the complex mobility
 \begin{equation}
  \label{eq:mu_salt_only}
  \mu (\omega) = \frac{ v_0 }{E_0} = - \frac{e}{\gamma} \frac{ 1 - i
    ( \frac{m \omega}{\gamma} ) }{ 1 + ( \frac{m\omega}{\gamma} )^2 }.
 \end{equation}
From the diffusion constant $D_I=0.60\,\sigma^2/\tau$ for the solution of salt concentration $\rho_s=0.0125\,\sigma^{-3}$, one obtains $\gamma=1.67\,m/\tau$. 
Eq.~(\ref{eq:mu_salt_only}) is plotted as dashed grey curves in Fig.~\ref{fig:7t2_cn}. 
The complex mobility of the counterions far away from the colloid ($r=8.0\,\sigma$) resembles that of free microions. 
The variation of mobility is characterized by a time scale set by
 \begin{equation}
  \tau_m = \frac{2\pi m}{\gamma}  \quad \text{or} \quad
  f_m = \frac{\gamma}{2 \pi m} .
 \end{equation}
Using the friction coefficient $\gamma=1.67\, m/\tau$, one obtains a frequency $f_m = 0.27 \, \tau^{-1}$. 
This is roughly the position of the maximum peak in the imaginary component of mobility for $r=8.0\,\sigma$. 
At frequencies higher than $f_m$, the finite mass of the counterions limits ions from responding to the external field, resulting in a reduction of the mobility.

\begin{figure}[htbp]
  \includegraphics[width=0.7\columnwidth]{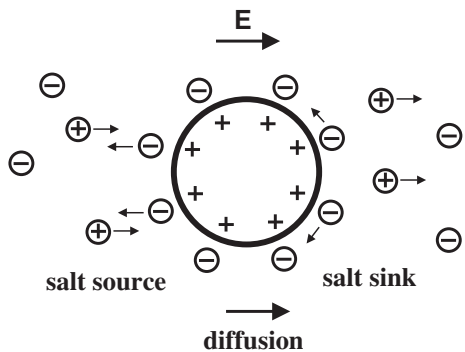}
  \caption{Scheme of the buildup of concentration gradient around a positive colloid. The salt concentration increases on the left-hand side of the colloid, and decreases on the right-hand side. As a consequence, a concentration-induced diffusion flux is formed in the direction of the external field. }
  \label{fig:salt}
\end{figure}

For counterions close to the colloid surface, their motion is
strongly influenced by the motion of the colloid. We shall examine
the mobility of counterions at distances $r=3.0\,\sigma$, starting
from low to high frequencies. At low frequency, one important
contribution comes from the enhanced salt concentration building up
near the colloid surface. The situation is depicted for a positively
charged colloid in Fig.~\ref{fig:salt}. At the back side of the
colloid, a wake of negative counterions builds up which are dragged
away from the surface by the external field. These counterions,
combined with the coions coming in from the bulk solution, create an
increase of the salt concentration on the left-hand side of the
colloid -- a salt source. A similar process occurs on the right-hand
side of the colloid, where depletion of the counterions and coions
results in a reduction of the salt concentration -- a salt sink.
Therefore, a salt concentration gradient is established along the
colloid surface, which drives the microions in the direction of the
external field. For counterions close to the surface, the effect due
to the concentration gradient, combined with the hydrodynamic drag
from the colloid, partially cancels out the effect of the external
field, which drives the counterions in the direction opposite to the
field. The result is an increase of the in-phase component of the
mobility at low frequency for counterions that are close to the
colloid surface.

The accumulation of the salt concentration cannot be accomplished
instantaneously, and the associated time scale is the time required
for microions to diffuse over the diameter of the colloid,
 \begin{equation}
  \tau_c = \frac{(2R)^2}{D_I}.
 \end{equation}
For our system, this time scale, $\tau_c=60\,\tau$, sets a frequency
scale,$f_c=1/\tau_c=0.017\,\tau^{-1}$. Above the crossover frequency
$f_c$, the variation of the external field is too fast for the
concentration gradient to build up, and the motion of the
counterions is mainly driven by the external field. Thus, the
response of the counterions reverts to that of free microions under
oscillating external field. At even higher frequency around $f_m$,
the inertia effect sets in and the mobility of the counterions is
reduced to zero.

\section{Dielectric response}
\label{sec:alpha}

\begin{figure}[htbp]
  \includegraphics[width=1.0\columnwidth]{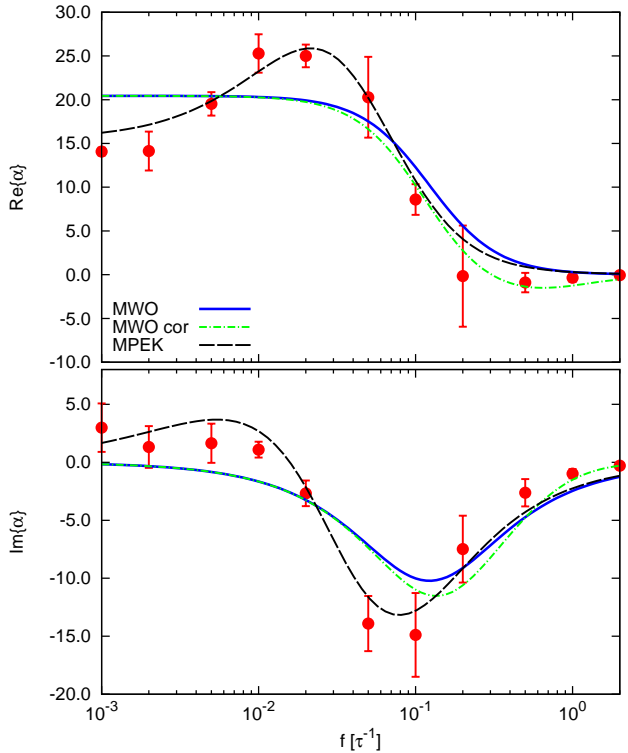}
  \caption{ Real and imaginary part of the complex polarizability $\alpha(\omega)$ of a charged colloid as a function of the frequency of applied electric field. The bare charge of the colloid is $Q=+50\,e$, and the salt concentration of the solution is $0.0125\,\sigma^{-3}$. The points with error bars are simulation data. The solid lines give the prediction from the
Maxwell-Wagner-O'Konski theory, and the dash-dotted lines show the revised prediction by taking into account of inertia effects. The dashed curves are numerical results from solving electrokinetic equations using the MPEK software. }
  \label{fig:7t2_alpha}
\end{figure}

In this section, we study the dielectric response of the charged
colloid in salt solution. The main quantity calculated from the
simulation is the complex polarizability $\alpha(\omega)$, which
characterizes the ratio between the dipole moment and the external
field. We focus on the effect of varying the frequency of the
external field. The amplitude of the field is chosen in the linear
region, $E_0 = 0.5 \, \varepsilon / (\sigma e)$, and frequency range
from $f=0.001$ to $2.0\, \tau^{-1}$ is considered. Fig.~\ref{fig:7t2_alpha} shows the real and imaginary parts of the
polarizability $\alpha(\omega)$ for a colloid particle with bare
charge $Q=+50\,e$. The solution has a salt concentration
$0.0125\,\sigma^{-3}$, corresponding to the Debye length $l_D =
1.72\,\sigma$ including 50 counterions. The Debye length is smaller
than, but comparable to the colloidal radius ($R=3.0\,\sigma$).

In the low-frequency region, the external perturbation is slow
enough that the system can follow, and the change of the
polarizability is connected to the polarization of the ionic cloud
surrounding the colloid. Two opposite effects take place at the same
time at low frequencies. One is due to the external field, which
drives the negative counterion cloud in the opposite direction of
the colloid movement. The resulting dipole moment points in the same
direction as the external field. The other effect is induced by the
salt concentration gradient near the colloid surface (see Fig.~\ref{fig:salt}), where counterions diffuse in the direction of the
external field. The concentration-induced effect results in a dipole
moment in the opposite direction of the field. The time required for
the concentration gradient to be established is $\tau_c= 60\,\tau$
(or $f_c = 0.017\, \tau^{-1}$). At frequencies higher than $f_c$,
the concentration-induced effect diminishes, resulting in an
increase of the in-phase component of the polarizability.

In the opposite limit of high frequency, the colloid and the ion
cloud can no longer follow the field; thus both ${\rm Re}\{\alpha\}$
and ${\rm Im}\{\alpha\}$ converge to zero. At intermediate
frequencies $f \sim 10^{-1}\,\tau^{-1}$, the real part ${\rm
Re}\{\alpha\}$ crosses over from positive to zero with an overshoot
below the transition frequency and a slight undershoot to negative
values after the transition frequency. The imaginary part ${\rm
Im}\{\alpha\}$ drops to large negative values, indicating that the
response is out of phase and that there is high dissipation. At high
frequencies, the main contribution to the dipole moment stems from
the conductivity mismatch between the colloid and the solvent due to
the presence of free counterions in the solvent. The finite time
required for the formation of the free charges is responsible for
the well-known Maxwell-Wagner dispersion, and this relaxation time
is given by
 \begin{equation}
  \label{eq:tmw}
  \tau_{\rm mw} = \frac{ \epsilon_p + 2 \epsilon_m }{ K_p + 2 K_m },
 \end{equation}
where $\epsilon$ and $K$ are the permittivity and conductivity of
the colloidal particle ($p$) and medium ($m$). For our simulation
model, $\epsilon_p=\epsilon_m$ and $K_p=0$. Using the relation
between the conductivity and the diffusion constant of the microion
(for 1-1 electrolytes), $K_m = 2 \rho_s e^2 D_I/k_B T$, the
Maxwell-Wagner relaxation time can be rewritten as
 \begin{equation}
  \tau_{\rm mw} = \frac{ 3\epsilon_m }{ 2 K_m } = \frac{3}{2} \frac{l_D^2}{D_I}.
 \end{equation}
Thus, $\tau_{\rm mw}$ is on the same order of the time required for
the microion to diffuse over the distance of Debye length. The
related frequency $f_{\rm mw}=1/\tau_{\rm mw}\simeq
0.12\,\tau^{-1}$. This is roughly the frequency at which ${\rm
Im}\{\alpha\}$ reaches a minimum.

We compare our simulation results with the predictions from two
theoretical models. One is the Maxwell-Wagner-O'Konski (MWO) theory
\cite{Saville2000}, which was originally developed for micrometer
colloids where the electric double layer is much thinner than the
colloid size ($\kappa R \gg 1$). We shall see that it still captures
some high-frequency features of the polarizability. We briefly
sketch the MWO theory in Appendix \ref{app:eMWO}. Another theory is
based on the standard electrokinetic model
\cite{DeLacey1981,Hill2003a}, which can be applied to arbitrary salt
concentrations. We summarize the governing equations of
electrokinetic model in Appendix \ref{app:ek}. In general, analytic
solutions only exist for large and small $\kappa R$ limits. For
intermediate value of $\kappa R$, one need to rely on numerical
methods to solve the electrokinetic equations. We compute the
complex polarizability using the software MPEK.

\begin{figure}[htbp]
  \includegraphics[width=1.0\columnwidth]{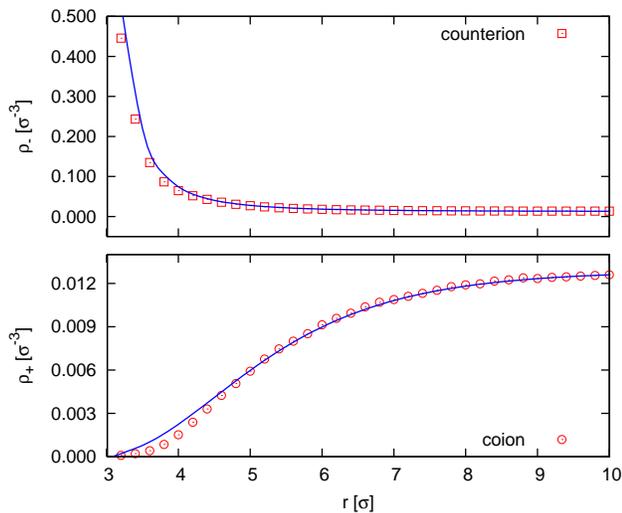}
  \caption{The densities for the negative counterions ($\rho_-$) and positive coions ($\rho_+$) as a function of the distance to the colloid center. The parameters are the same as in Fig.~\ref{fig:7t2_alpha}. The lines are results by solving the Poisson-Boltzmann equation \cite{Baptista2009,phd_Schmitz}.}
  \label{fig:7t2_ion}
\end{figure}

One important quantity required as an input for the theory is the
zeta potential, defined as the electric potential at the shear
plane, an imaginary plane separating the hydrodynamically mobile and
immobile fluid. The exact position of the shear plane is difficult
to determine, and it is also possible that there is no sharp
boundary at all. Since the interacting sites of the colloid are set
at $R=3.0\,\sigma$, and in our model, they are responsible for the
hydrodynamical coupling to the fluid, it is reasonable to use
$R=3.0\,\sigma$ for the position of the shear plane. The
zeta-potential can be obtained by solving the Poisson-Boltzmann
equation. For a spherical particle, numerical tables for the
solution to Poisson-Boltzmann equation were given by Loeb \emph{et
al.} \cite{LOW_table}. An analytic expression for the relationship
between the $\zeta$-potential and the surface charge density
$\sigma$ was derived by Ohshima \emph{et al.} \cite{Ohshima1982,
Ohshima}
 \begin{eqnarray}
  \label{eq:Ohshima}
  \sigma &=& \frac{2\epsilon_m \kappa k_B T}{e} \sinh \left( \frac{e\zeta}{2k_BT} \right) \bigg[ 1 + \frac{1}{\kappa R} \frac{2}{\cosh^2(e\zeta/4k_BT)} \nonumber \\
  && + \frac{1}{(\kappa R)^2} \frac{ 8\ln [ \cosh(e\zeta/4k_BT)] }{ \sinh^2(e\zeta/2k_BT) } \bigg]^{1/2}
 \end{eqnarray}
For our system, we obtain a scaled zeta potential
$\zeta=4.12\,\varepsilon/e$. We also measure the charge density as a
function of the distance to the colloid center. The simulation
results are shown in Fig.~\ref{fig:7t2_ion}. Also shown are the
numerical solutions from solving the Poisson-Boltzmann equation by a
variational approach \cite{Baptista2009,phd_Schmitz}. The simulation
and numerical results show good agreement. By integrating the charge
density from the simulation, we obtain the zeta potential
$\zeta=4.26\,\varepsilon/e$, close to the value computed using
Ohshima's formula (\ref{eq:Ohshima}). We use the simulation value in
the following calculation.

According to the Maxwell-Wagner-O'Konski theory, the effect of the electric double layer can be taken into account by introducing a surface conductance in the Maxwell-Wagner theory. 
We estimate the surface conductance, $K_{\sigma}=0.23\,e^2/(\sigma\varepsilon\tau)$, based on Bikerman's expression (see Appendix \ref{app:eMWO} for details). 
Using this value, we can calculate the theoretical polarizability as a function of frequency without any fitting parameters. 
The resulting curves are shown in Figure \ref{fig:7t2_alpha} with solid lines. The result of the MWO theory is only in qualitative agreement with the simulation. 
The theory captures the main qualitative features, and roughly the correct crossover frequency, but it misses most of the details around the transition frequency.

The simulation results feature a slight undershoot at the high frequency region. 
This can be explained by considering the inertia effect of the microions, which are not entirely negligible in our simulations. (In real systems, they are negligible.) 
The inertia effect can be incorporated into the theory by a frequency-dependent
conductivity (see Appendix \ref{app:eMWO}) and the theory can be revised accordingly, giving the dash-dotted lines in Figure \ref{fig:7t2_alpha}. 
It predicts small undershoots in both the real and the imaginary part of the polarizability, which are consistent with the simulation. 
However, both results from the MWO theory, the original one and the one with the inertia correction, miss the overshoot of the ${\rm Re}\{\alpha\}$ at the low-frequency regime.

In the Maxwell-Wagner theory, the colloid and its surrounding medium
are treated as homogeneous substances, and only the bulk properties
of the components are taken into account. The induced charges due to
the external field only appear at the colloid-fluid interface, and
the effect of the spatial distribution of the polarization charges
is neglected. This simplification is valid for large colloids, but
for small particles ($\kappa R \sim 1$), the distribution of the
polarization charges near the interface may become important. The
theory taking into consideration of the effect of space charge
variation is based on standard electrokinetic equations. One
important result from the electrokinetic theory is the low-frequency
dielectric dispersion \cite{DukhinShilov}. We use the program MPEK
to calculate the complex polarizability, and the results are shown
as dashed lines in Fig.~\ref{fig:7t2_alpha}. The prediction from the
electrokinetic theory is in good agreement with the simulation, and
correctly captures the overshoot at low frequency.

\section{Summary}
\label{sec:summary}

We have carried out mesoscopic molecular dynamics simulations for a charged colloidal particle under static and alternating electric fields, with different colloid charges and salt concentrations. 
We have fully taken into account the hydrodynamic interaction with thermal fluctuations, using Dissipative Particle Dynamics, and the electrostatic interactions, using the Particle-Particle-Particle Mesh method.

The motion of a charged colloidal particle under the influence of external fields is a complicated process due to the presence of the electric double layer. 
We have investigated the mobility of both the colloid and its surrounding ionic cloud. 
For free microions, their motion is simple: cations move in the direction of the field, and anions move in the opposite direction. 
However, due to the presence of the colloid, the microion motion is greatly modified, especially for counterions close to the colloid surface. 
The hydrodynamic and electrostatic interactions with the colloid are then strong enough to force them to move together with the colloid. 
The range of the interaction from the colloid can be tuned by varying the salt
concentration or the colloid charges. 
In general, the range is short for high ionic strength or weakly charged colloids.

When an alternating electric field is applied, the dynamic response of the colloid and micrions depends on the frequency. 
The frequency-dependence of the microion mobility is controlled by two different processes. 
On the one hand, the oscillating field drives the microions directly. At low frequency, the movement of the microion is in phase with the perturbation. 
At high frequency, the inertia effect prevents the microion from following the external field. 
The frequency that separates these two different region is related to the time scale $\tau_m$ . 
On the other hand, a concentration gradient can build up around the colloid surface at low frequency, which drives a diffusive motion of the microions from
the region of high concentration to that of low concentration. 
The time required for the gradient to develop is $\tau_c$. 
The motion of the microion is determined by the interplay between the field-induced and concentration-induced effects. 
For microions away from the colloid surface, the interaction to the colloid is weak, and the field-induced effect dominates. 
For microions close to the surface, the microions follow the colloid's motion at low frequency, but recover to the free-ion case at high frequency.

The dielectric response of the charged colloid is characterized by the polarizability. For our colloid model ($\epsilon_p=\epsilon_m$ and $K_p=0$), the main contribution to the dipole moment comes from the electric double layer. 
For microions, there are two important length scales: one is the Debye screening length $l_D$, and the other one is the diameter of the colloid $2R$. 
Associated with these two length scales, are the two time scales $\tau_{\rm mw}$ and $\tau_c$, which correspond to the time required for the microion to diffuse over the distance $l_D$ and $2R$, respectively. 
For the parameters considered in this work, $2R>l_D$, thus the frequency $f_{\rm mw}$ is larger than $f_c$. 
The competition of these two time scales results in a maximum of ${\rm Re}\{\alpha\}$ at the intermediate frequency $f_c<f<f_{\rm mw}$. 
We compare the simulation results to the predictions from the Maxwell-Wagner-O'Konski theory and standard electrokinetic model.
The MWO theory only captures the dielectric response in the high-frequency region, while the electrokinetic theory performs well over the whole frequency range.

\begin{acknowledgments}
We are grateful to Prof. Reghan Hill of McGill University for providing the computer program MPEK. 
We thank Peter Virnau and Stefan Medina Hernando for discussions. 
This work is funded by the Deutsche Forschungsgemeinschaft (DFG) through the SFB-TR6 program ``Physics of Colloidal Dispersions in External Fields''.
Computational resources at John von Neumann Institute for Computing (NIC J\"ulich), High Performance Computing Center Stuttgart (HLRS) and JGU Mainz (MOGON) are gratefully acknowledged.
\end{acknowledgments}

\appendix

\section{Maxwell-Wagner-O'Konski theory}
\label{app:eMWO}

In this appendix, we give a short introduction to the Maxwell-Wagner-O'Konski theory.
More detailed information can be found in Refs. \cite{Maxwell1954, Wagner1914, OKonski1960, Saville2000}.

Assume an external electric field with a form ${\bf E} = E_0 e^{i\omega t} \hat{\bf x}$ is applied to a colloidal solution.
The external field induces a dipole moment to the colloidal particle.
The amplitude of the dipole moment can be written as $ p_0 = 4 \pi \epsilon_m K(\omega) R^3 E_0$ (cf. Equation (\ref{eq:alpha_def})), where the Clausius-Mossotti factor $K(\omega)$ is a complex number containing both the magnitude and the phase information.
In the Maxwell-Wagner theory, it has the form
\begin{equation}
  \label{eq:cm2}
  K (\epsilon^*_p, \epsilon^*_m)
    = \frac{ \epsilon^*_p - \epsilon^*_m }{ \epsilon^*_p + 2\epsilon^*_m },
\end{equation}
where $\epsilon^*_p$ and $\epsilon^*_m$ are the complex dielectric constants of the particle and the medium, respectively.
They are defined as
\begin{equation}
  \epsilon^*_p = \epsilon_p + \frac{ K_p }{i\omega}, \quad
  \epsilon^*_m = \epsilon_m + \frac{ K_m }{i\omega},
\end{equation}
where $\epsilon$ (without the star) and $K$ are the permittivity and conductivity, respectively.
For $\epsilon_m=\epsilon_p=\epsilon$, the Clausius-Mossotti factor can be rewritten as
\begin{equation}
\label{eq:K}
K =K_0 \: \frac{1-i \tilde{\omega}}{1+\tilde{\omega}^2}
\end{equation}
with
\begin{equation}
K_0 = \frac{K_p - K_m}{K_p + 2 K_m} \quad \mbox{and} \quad
\tilde{\omega} = \frac{3 \epsilon}{K_p + 2 K_m} \: \omega.
\end{equation}
The conductivity of the medium $K_m$ is related to the microion diffusion constant $D_I$  {\em via} $K_m = 2 \rho_s e^2 D_I / (k_BT)$ for 1-1 electrolytes.

The classical Maxwell-Wagner theory fails to explain the dielectrophoretic properties of latex particles.
Latex has a low intrinsic conductivity, but the measurement indicated that the particle conductivity is high.
It was concluded that the electric double layer surface significantly contributes to the particle conductivity.
This was first demonstrated by O'Konski
\cite{OKonski1960}
\begin{equation}
  \label{eq:OKonski}
  K_p \rightarrow K_p + \frac {2K_{\sigma}}{R}
\end{equation}
where $K_{\sigma}$ is the surface conductance (unit S instead of S$\cdot$m$^{-1}$ for conductivity) due to the electric double layer.

The surface conductance then can be related to the $\zeta$-potential by Bikerman's expression
\begin{equation}
  \label{eq:Bikerman}
  K_{\sigma} = l_D \left[ \exp (| \frac{e \zeta}{2k_BT} |) -1 \right] (1+3m) K_m,
\end{equation}
where $m$ is a dimensionless ionic drag coefficient \cite{Bikerman1940,OBrien1986}.
Bikerman's expression is applicable for 1-1 electrolytes, and the contribution from coions has been neglected.
The ionic drag coefficient is related to the viscosity $\eta_s$ and ion diffusion constant $D_I$,
\begin{equation}
  m = \frac{ 2\epsilon_m (k_B T)^2 } { 3 \eta_s e^2 D_I },
\end{equation}
($m=0.072$ for salt density $\rho_s=0.0125\,\sigma^{-3}$, and $m=0.18$ for KCl).
With $Q=50e$ and $R=3.0\sigma$, we obtain the surface conductance $K_{\sigma}=0.23\, e^2/(\sigma\sqrt{m\varepsilon})$.

The inertia effect of ions can be taken into account by considering a charged particle with charge $e$, mass $m$ immersed in a viscous fluid.
The equation of motion for the particle under an AC field $E_0 e^{i\omega t}$ is
\begin{equation}
  m \ddot{x} = - \gamma \dot{x} + eE_0 e^{i\omega t},
\end{equation}
where $\gamma = k_BT/D_I$ is the friction constant.
After solving the equation of motion, one finds that the velocity of the particle has the form
\begin{equation}
  \label{eq:v_omega}
  \dot{x} \propto \frac{ 1 - i ( \frac{m\omega}{\gamma} )}{ 1 +
    (\frac{m\omega}{\gamma})^2}.
\end{equation}
Since the conductivity of salt solutions is proportional to the ion's velocity, it is reasonable to assume that the conductivity has the same frequency dependency
\begin{equation}
  \label{eq:Km_omega}
  K_m \rightarrow K_m \frac{ 1 - i ( \frac{m\omega}{\gamma} )}{ 1 +
    (\frac{m\omega}{\gamma})^2}.
\end{equation}
The corrected MWO results in Fig.~\ref{fig:7t2_alpha} are obtained by substituting (\ref{eq:Km_omega}) into the Clausius-Mossotti factor (\ref{eq:K}).

\section{Electrokinetic equations}
\label{app:ek}

Consider a colloidal particle immersed in an electrolyte solution.
The solution consists of $N$ ionic species of charge $z_ke$ and bulk concentration $c_k(\infty)$.
The spherical colloid has a hydrodynamic radius of $R$.
The external electric field is $\mathbf{E}$.

The system is described in terms of the electrostatic potential
$\psi(\mathbf{r})$, the ion concentration field $c_k(\mathbf{r})$,
the flow velocity field $\mathbf{u}(\mathbf{r})$ and the pressure
$p(\mathbf{r})$. The governing laws of the system are the Poisson
equation, the Nernst-Planck equation and the Navier-Stokes
equations.

\emph{Poisson Equation}

The electrostatic potential $\psi$ outside the colloid is related to
the ion concentration by the Poisson equation
\begin{equation}
  \nabla^2 \psi = -\frac{1}{\epsilon_m} \sum_k z_k e c_k,
\end{equation}
where $\epsilon_m$ is the solvent permittivity.
The boundary conditions for the potential are
\begin{eqnarray}
  \phi=\zeta, && r=R, \\
  \phi=-\mathbf{E}\cdot\mathbf{r}, && r\rightarrow \infty.
\end{eqnarray}

\emph{Nernst-Planck Equation}

The flux of type-$k$ ion can be written as
\begin{equation}
\label{eq:ion_flux}
  \mathbf{j}_k = -D_k \nabla c_k - z_k D_k ( \nabla \frac{e\psi}{k_BT} ) c_k + \mathbf{u} c_k.
\end{equation}
The first term is the contribution from the ion's diffusion due to the concentration gradient.
The second term represents the effect of electric field.
The last term expresses the flow-induced hydrodynamic drag force.

The conservation of ion number is described by a continuity equation,
\begin{equation}
\label{eq:ion_continuity}
  \frac{\partial c_k}{\partial t} + \nabla \cdot \mathbf{j}_k =0,
\end{equation}
After substituting Eq.~(\ref{eq:ion_flux}) and omitting the
subscript $k$ for simplicity, Eq.~(\ref{eq:ion_continuity}) takes
the form
\begin{equation}
  \frac{\partial c}{\partial t} - \nabla \cdot (D \nabla c)
  - \nabla \cdot \left( z D (\nabla \frac{e\psi}{k_BT}) c \right)
  + \nabla \cdot (\mathbf{u}c) =0.
\end{equation}
The boundary condition at the colloid surface is
\begin{equation}
  \mathbf{j} \cdot \mathbf{n} = 0, \quad  r=R,
\end{equation}
where $\mathbf{n}$ is the normal to the surface.
Far away from the colloid, the concentration reaches to the bulk value
\begin{equation}
  c=c(\infty). \quad r \rightarrow \infty.
\end{equation}

\emph{Navier-Stokes Equation}

The Navier-Stokes equation with an external body force is
\begin{equation}
  \rho_0 \frac{\partial \mathbf{u}}{\partial t}
  + \rho_0 \mathbf{u} \cdot \nabla \mathbf{u}
  = \eta_s \nabla^2 \mathbf{u} -\nabla p -\nabla \psi \sum_k z_k e c_k,
\end{equation}
where $\rho_0$ is the solvent density and the last term on the right-hand side is due to the electric field.
The second term on the left-hand side can be dropped for incompressible fluid, $\nabla \cdot \mathbf{u}=0$.
No-slip boundary condition at the colloid surface requires
\begin{equation}
  \mathbf{u} = \mathbf{V}, \quad r=R,
\end{equation}
where $\mathbf{V}$ is the velocity of the colloid.


\begin{thebibliography}{57}
\expandafter\ifx\csname natexlab\endcsname\relax\def\natexlab#1{#1}\fi
\expandafter\ifx\csname bibnamefont\endcsname\relax
  \def\bibnamefont#1{#1}\fi
\expandafter\ifx\csname bibfnamefont\endcsname\relax
  \def\bibfnamefont#1{#1}\fi
\expandafter\ifx\csname citenamefont\endcsname\relax
  \def\citenamefont#1{#1}\fi
\expandafter\ifx\csname url\endcsname\relax
  \def\url#1{\texttt{#1}}\fi
\expandafter\ifx\csname urlprefix\endcsname\relax\def\urlprefix{URL }\fi
\providecommand{\bibinfo}[2]{#2}
\providecommand{\eprint}[2][]{\url{#2}}

\bibitem[{\citenamefont{Russel et~al.}(1989)\citenamefont{Russel, Saville, and
  Schowalter}}]{RSS}
\bibinfo{author}{\bibfnamefont{W.~B.} \bibnamefont{Russel}},
  \bibinfo{author}{\bibfnamefont{D.~A.} \bibnamefont{Saville}},
  \bibnamefont{and}
  \bibinfo{author}{\bibfnamefont{W.}~\bibnamefont{Schowalter}},
  \emph{\bibinfo{title}{Colloidal Dispersions}} (\bibinfo{publisher}{Cambridge
  University Press}, \bibinfo{address}{Cambridge}, \bibinfo{year}{1989}).

\bibitem[{\citenamefont{Hiemenz and Rajagopalan}(1997)}]{Hiemenz_colloid3}
\bibinfo{author}{\bibfnamefont{P.~C.} \bibnamefont{Hiemenz}} \bibnamefont{and}
  \bibinfo{author}{\bibfnamefont{R.}~\bibnamefont{Rajagopalan}},
  \emph{\bibinfo{title}{Principles of Colloid and Surface Chemistry}}
  (\bibinfo{publisher}{Marcel Dekker}, \bibinfo{address}{New York},
  \bibinfo{year}{1997}), \bibinfo{edition}{3rd} ed.

\bibitem[{\citenamefont{Dhont}(1996)}]{Dhont}
\bibinfo{author}{\bibfnamefont{J.}~\bibnamefont{Dhont}},
  \emph{\bibinfo{title}{An Introduction to Dynamics of Colloids}}
  (\bibinfo{publisher}{Elsevier}, \bibinfo{address}{Amsterdam},
  \bibinfo{year}{1996}).

\bibitem[{\citenamefont{Fischer et~al.}(2008)\citenamefont{Fischer, Naji, and
  Netz}}]{Fischer2008}
\bibinfo{author}{\bibfnamefont{S.}~\bibnamefont{Fischer}},
  \bibinfo{author}{\bibfnamefont{A.}~\bibnamefont{Naji}}, \bibnamefont{and}
  \bibinfo{author}{\bibfnamefont{R.~R.} \bibnamefont{Netz}},
  \bibinfo{journal}{Phys. Rev. Lett.} \textbf{\bibinfo{volume}{101}},
  \bibinfo{pages}{176103} (\bibinfo{year}{2008}).

\bibitem[{\citenamefont{Dhont and Kang}(2010)}]{Dhont2010}
\bibinfo{author}{\bibfnamefont{J.}~\bibnamefont{Dhont}} \bibnamefont{and}
  \bibinfo{author}{\bibfnamefont{K.}~\bibnamefont{Kang}},
  \bibinfo{journal}{Eur. Phys. J. E} \textbf{\bibinfo{volume}{33}},
  \bibinfo{pages}{51} (\bibinfo{year}{2010}).

\bibitem[{\citenamefont{Zhou and Schmid}(2013)}]{2012_q0}
\bibinfo{author}{\bibfnamefont{J.}~\bibnamefont{Zhou}} \bibnamefont{and}
  \bibinfo{author}{\bibfnamefont{F.}~\bibnamefont{Schmid}},
  \bibinfo{journal}{Eur. Phys. J. E} \textbf{\bibinfo{volume}{36}},
  \bibinfo{pages}{33} (\bibinfo{year}{2013}).

\bibitem[{\citenamefont{Maxwell}(1954)}]{Maxwell1954}
\bibinfo{author}{\bibfnamefont{J.}~\bibnamefont{Maxwell}},
  \emph{\bibinfo{title}{Electricity and Magnetism, vol. 1}}
  (\bibinfo{publisher}{Dover}, \bibinfo{address}{New York},
  \bibinfo{year}{1954}).

\bibitem[{\citenamefont{Wagner}(1914)}]{Wagner1914}
\bibinfo{author}{\bibfnamefont{K.}~\bibnamefont{Wagner}},
  \bibinfo{journal}{Arch. Electrotech} \textbf{\bibinfo{volume}{2}},
  \bibinfo{pages}{371} (\bibinfo{year}{1914}).

\bibitem[{\citenamefont{Green and Morgan}(1999)}]{Green1999}
\bibinfo{author}{\bibfnamefont{N.~G.} \bibnamefont{Green}} \bibnamefont{and}
  \bibinfo{author}{\bibfnamefont{H.}~\bibnamefont{Morgan}},
  \bibinfo{journal}{J. Phys. Chem. B} \textbf{\bibinfo{volume}{103}},
  \bibinfo{pages}{41} (\bibinfo{year}{1999}).

\bibitem[{\citenamefont{Ermolina and Morgan}(2005)}]{Ermolina2005}
\bibinfo{author}{\bibfnamefont{I.}~\bibnamefont{Ermolina}} \bibnamefont{and}
  \bibinfo{author}{\bibfnamefont{H.}~\bibnamefont{Morgan}},
  \bibinfo{journal}{J. Colloid Interface Sci.} \textbf{\bibinfo{volume}{285}},
  \bibinfo{pages}{419} (\bibinfo{year}{2005}).

\bibitem[{\citenamefont{O'Konski}(1960)}]{OKonski1960}
\bibinfo{author}{\bibfnamefont{C.}~\bibnamefont{O'Konski}},
  \bibinfo{journal}{J. Phys. Chem.} \textbf{\bibinfo{volume}{64}},
  \bibinfo{pages}{605} (\bibinfo{year}{1960}).

\bibitem[{\citenamefont{Saville et~al.}(2000)\citenamefont{Saville, Bellini,
  Degiorgio, and Mantegazza}}]{Saville2000}
\bibinfo{author}{\bibfnamefont{D.~A.} \bibnamefont{Saville}},
  \bibinfo{author}{\bibfnamefont{T.}~\bibnamefont{Bellini}},
  \bibinfo{author}{\bibfnamefont{V.}~\bibnamefont{Degiorgio}},
  \bibnamefont{and}
  \bibinfo{author}{\bibfnamefont{F.}~\bibnamefont{Mantegazza}},
  \bibinfo{journal}{J. Chem. Phys.} \textbf{\bibinfo{volume}{113}},
  \bibinfo{pages}{6974} (\bibinfo{year}{2000}).

\bibitem[{\citenamefont{Dukhin and Shilov}(1974)}]{DukhinShilov}
\bibinfo{author}{\bibfnamefont{S.}~\bibnamefont{Dukhin}} \bibnamefont{and}
  \bibinfo{author}{\bibfnamefont{V.}~\bibnamefont{Shilov}},
  \emph{\bibinfo{title}{Dielectric phenomena and the double layer in disperse
  systems and polyelectrolytes}} (\bibinfo{publisher}{Wiley},
  \bibinfo{address}{New York}, \bibinfo{year}{1974}).

\bibitem[{\citenamefont{O'Brien and White}(1978)}]{OBrien1978}
\bibinfo{author}{\bibfnamefont{R.~W.} \bibnamefont{O'Brien}} \bibnamefont{and}
  \bibinfo{author}{\bibfnamefont{L.~R.} \bibnamefont{White}},
  \bibinfo{journal}{J. Chem. Soc.{,} Faraday Trans. 2}
  \textbf{\bibinfo{volume}{74}}, \bibinfo{pages}{1607} (\bibinfo{year}{1978}).

\bibitem[{\citenamefont{DeLacey and White}(1981)}]{DeLacey1981}
\bibinfo{author}{\bibfnamefont{E.~H.~B.} \bibnamefont{DeLacey}}
  \bibnamefont{and} \bibinfo{author}{\bibfnamefont{L.~R.} \bibnamefont{White}},
  \bibinfo{journal}{J. Chem. Soc.{,} Faraday Trans. 2}
  \textbf{\bibinfo{volume}{77}}, \bibinfo{pages}{2007} (\bibinfo{year}{1981}).

\bibitem[{\citenamefont{Hill et~al.}(2003{\natexlab{a}})\citenamefont{Hill,
  Saville, and Russel}}]{Hill2003}
\bibinfo{author}{\bibfnamefont{R.~J.} \bibnamefont{Hill}},
  \bibinfo{author}{\bibfnamefont{D.~A.} \bibnamefont{Saville}},
  \bibnamefont{and} \bibinfo{author}{\bibfnamefont{W.~B.}
  \bibnamefont{Russel}}, \bibinfo{journal}{Phys. Chem. Chem. Phys.}
  \textbf{\bibinfo{volume}{5}}, \bibinfo{pages}{911}
  (\bibinfo{year}{2003}{\natexlab{a}}).

\bibitem[{\citenamefont{Kim et~al.}(2006)\citenamefont{Kim, Nakayama, and
  Yamamoto}}]{Kim2006}
\bibinfo{author}{\bibfnamefont{K.}~\bibnamefont{Kim}},
  \bibinfo{author}{\bibfnamefont{Y.}~\bibnamefont{Nakayama}}, \bibnamefont{and}
  \bibinfo{author}{\bibfnamefont{R.}~\bibnamefont{Yamamoto}},
  \bibinfo{journal}{Phys. Rev. Lett.} \textbf{\bibinfo{volume}{96}},
  \bibinfo{pages}{208306} (\bibinfo{year}{2006}).

\bibitem[{\citenamefont{Y.~Nakayama and Yamamoto}(2008)}]{Nakayama2008}
\bibinfo{author}{\bibfnamefont{K.~K.} \bibnamefont{Y.~Nakayama}}
  \bibnamefont{and} \bibinfo{author}{\bibfnamefont{R.}~\bibnamefont{Yamamoto}},
  \bibinfo{journal}{Eur. Phys. J. E} \textbf{\bibinfo{volume}{26}},
  \bibinfo{pages}{361} (\bibinfo{year}{2008}).

\bibitem[{\citenamefont{Zhao and Bau}(2009)}]{Zhaohui2009}
\bibinfo{author}{\bibfnamefont{H.}~\bibnamefont{Zhao}} \bibnamefont{and}
  \bibinfo{author}{\bibfnamefont{H.~H.} \bibnamefont{Bau}},
  \bibinfo{journal}{J. Colloid Interface Sci.} \textbf{\bibinfo{volume}{333}},
  \bibinfo{pages}{663} (\bibinfo{year}{2009}).

\bibitem[{\citenamefont{Schmitz and D\"unweg}(2012)}]{Schmitz2012}
\bibinfo{author}{\bibfnamefont{R.}~\bibnamefont{Schmitz}} \bibnamefont{and}
  \bibinfo{author}{\bibfnamefont{B.}~\bibnamefont{D\"unweg}},
  \bibinfo{journal}{J. Phys.: Condens. Matter} \textbf{\bibinfo{volume}{24}},
  \bibinfo{pages}{464111} (\bibinfo{year}{2012}).

\bibitem[{\citenamefont{Liu et~al.}(2010)\citenamefont{Liu, Zhu, and
  Maginn}}]{LiuHongjun2010}
\bibinfo{author}{\bibfnamefont{H.}~\bibnamefont{Liu}},
  \bibinfo{author}{\bibfnamefont{Y.}~\bibnamefont{Zhu}}, \bibnamefont{and}
  \bibinfo{author}{\bibfnamefont{E.}~\bibnamefont{Maginn}},
  \bibinfo{journal}{Macromolecules} \textbf{\bibinfo{volume}{43}},
  \bibinfo{pages}{4805} (\bibinfo{year}{2010}).

\bibitem[{\citenamefont{Hsiao et~al.}(2011)\citenamefont{Hsiao, Wei, and
  Chang}}]{HsiaoPai-Yi2011}
\bibinfo{author}{\bibfnamefont{P.-Y.} \bibnamefont{Hsiao}},
  \bibinfo{author}{\bibfnamefont{Y.-F.} \bibnamefont{Wei}}, \bibnamefont{and}
  \bibinfo{author}{\bibfnamefont{H.-C.} \bibnamefont{Chang}},
  \bibinfo{journal}{Soft Matter} \textbf{\bibinfo{volume}{7}},
  \bibinfo{pages}{1207} (\bibinfo{year}{2011}).

\bibitem[{\citenamefont{Zhang et~al.}(2012)\citenamefont{Zhang, Xiang, and
  Hu}}]{ZhangQi-Yi2012}
\bibinfo{author}{\bibfnamefont{Q.-Y.} \bibnamefont{Zhang}},
  \bibinfo{author}{\bibfnamefont{X.}~\bibnamefont{Xiang}}, \bibnamefont{and}
  \bibinfo{author}{\bibfnamefont{K.-Y.} \bibnamefont{Hu}},
  \bibinfo{journal}{Modern Physics Letters B} \textbf{\bibinfo{volume}{26}},
  \bibinfo{pages}{1250089} (\bibinfo{year}{2012}).

\bibitem[{\citenamefont{Sigalov et~al.}(2008)\citenamefont{Sigalov, Comer,
  Timp, and Aksimentiev}}]{Sigalov2008}
\bibinfo{author}{\bibfnamefont{G.}~\bibnamefont{Sigalov}},
  \bibinfo{author}{\bibfnamefont{J.}~\bibnamefont{Comer}},
  \bibinfo{author}{\bibfnamefont{G.}~\bibnamefont{Timp}}, \bibnamefont{and}
  \bibinfo{author}{\bibfnamefont{A.}~\bibnamefont{Aksimentiev}},
  \bibinfo{journal}{Nano Lett.} \textbf{\bibinfo{volume}{8}},
  \bibinfo{pages}{56} (\bibinfo{year}{2008}).

\bibitem[{\citenamefont{Lobaskin and D\"unweg}(2004)}]{Lobaskin2004}
\bibinfo{author}{\bibfnamefont{V.}~\bibnamefont{Lobaskin}} \bibnamefont{and}
  \bibinfo{author}{\bibfnamefont{B.}~\bibnamefont{D\"unweg}},
  \bibinfo{journal}{New Journal of Physics} \textbf{\bibinfo{volume}{6}},
  \bibinfo{pages}{54} (\bibinfo{year}{2004}).

\bibitem[{\citenamefont{Lobaskin et~al.}(2004)\citenamefont{Lobaskin, D\"unweg,
  and Holm}}]{Lobaskin2004a}
\bibinfo{author}{\bibfnamefont{V.}~\bibnamefont{Lobaskin}},
  \bibinfo{author}{\bibfnamefont{B.}~\bibnamefont{D\"unweg}}, \bibnamefont{and}
  \bibinfo{author}{\bibfnamefont{C.}~\bibnamefont{Holm}}, \bibinfo{journal}{J.
  Phys.: Condens. Matter} \textbf{\bibinfo{volume}{16}}, \bibinfo{pages}{S4063}
  (\bibinfo{year}{2004}).

\bibitem[{\citenamefont{Lobaskin et~al.}(2007)\citenamefont{Lobaskin, D\"unweg,
  Medebach, Palberg, and Holm}}]{Lobaskin2007}
\bibinfo{author}{\bibfnamefont{V.}~\bibnamefont{Lobaskin}},
  \bibinfo{author}{\bibfnamefont{B.}~\bibnamefont{D\"unweg}},
  \bibinfo{author}{\bibfnamefont{M.}~\bibnamefont{Medebach}},
  \bibinfo{author}{\bibfnamefont{T.}~\bibnamefont{Palberg}}, \bibnamefont{and}
  \bibinfo{author}{\bibfnamefont{C.}~\bibnamefont{Holm}},
  \bibinfo{journal}{Phys. Rev. Lett.} \textbf{\bibinfo{volume}{98}},
  \bibinfo{pages}{176105} (\bibinfo{year}{2007}).

\bibitem[{\citenamefont{Chatterji and Horbach}(2005)}]{Chatterji2005}
\bibinfo{author}{\bibfnamefont{A.}~\bibnamefont{Chatterji}} \bibnamefont{and}
  \bibinfo{author}{\bibfnamefont{J.}~\bibnamefont{Horbach}},
  \bibinfo{journal}{J. Chem. Phys.} \textbf{\bibinfo{volume}{122}},
  \bibinfo{eid}{184903} (\bibinfo{year}{2005}).

\bibitem[{\citenamefont{Chatterji and Horbach}(2007)}]{Chatterji2007}
\bibinfo{author}{\bibfnamefont{A.}~\bibnamefont{Chatterji}} \bibnamefont{and}
  \bibinfo{author}{\bibfnamefont{J.}~\bibnamefont{Horbach}},
  \bibinfo{journal}{J. Chem. Phys.} \textbf{\bibinfo{volume}{126}},
  \bibinfo{eid}{064907} (\bibinfo{year}{2007}).

\bibitem[{\citenamefont{Giupponi and Pagonabarraga}(2011)}]{Giupponi2011}
\bibinfo{author}{\bibfnamefont{G.}~\bibnamefont{Giupponi}} \bibnamefont{and}
  \bibinfo{author}{\bibfnamefont{I.}~\bibnamefont{Pagonabarraga}},
  \bibinfo{journal}{Phys. Rev. Lett.} \textbf{\bibinfo{volume}{106}},
  \bibinfo{pages}{248304} (\bibinfo{year}{2011}).

\bibitem[{\citenamefont{Malevanets and Kapral}(1999)}]{Malevanets1999}
\bibinfo{author}{\bibfnamefont{A.}~\bibnamefont{Malevanets}} \bibnamefont{and}
  \bibinfo{author}{\bibfnamefont{R.}~\bibnamefont{Kapral}},
  \bibinfo{journal}{J. Chem. Phys.} \textbf{\bibinfo{volume}{110}},
  \bibinfo{pages}{8605} (\bibinfo{year}{1999}).

\bibitem[{\citenamefont{Gompper et~al.}(2009)\citenamefont{Gompper, Ihle,
  Kroll, and Winkler}}]{Gompper2009}
\bibinfo{author}{\bibfnamefont{G.}~\bibnamefont{Gompper}},
  \bibinfo{author}{\bibfnamefont{T.}~\bibnamefont{Ihle}},
  \bibinfo{author}{\bibfnamefont{D.~M.} \bibnamefont{Kroll}}, \bibnamefont{and}
  \bibinfo{author}{\bibfnamefont{R.~G.} \bibnamefont{Winkler}},
  \bibinfo{journal}{Adv. Polym. Sci.} \textbf{\bibinfo{volume}{221}},
  \bibinfo{pages}{1} (\bibinfo{year}{2009}).

\bibitem[{\citenamefont{Hoogerbrugge and Koelman}(1992)}]{Hoogerbrugge1992}
\bibinfo{author}{\bibfnamefont{P.~J.} \bibnamefont{Hoogerbrugge}}
  \bibnamefont{and} \bibinfo{author}{\bibfnamefont{J.~M. V.~A.}
  \bibnamefont{Koelman}}, \bibinfo{journal}{Europhys. Lett.}
  \textbf{\bibinfo{volume}{19}}, \bibinfo{pages}{155} (\bibinfo{year}{1992}).

\bibitem[{\citenamefont{Espa\~nol and Warren}(1995)}]{Espanol1995}
\bibinfo{author}{\bibfnamefont{P.}~\bibnamefont{Espa\~nol}} \bibnamefont{and}
  \bibinfo{author}{\bibfnamefont{P.~B.} \bibnamefont{Warren}},
  \bibinfo{journal}{Europhys. Lett.} \textbf{\bibinfo{volume}{30}},
  \bibinfo{pages}{191} (\bibinfo{year}{1995}).

\bibitem[{\citenamefont{Groot and Warren}(1997)}]{Groot1997}
\bibinfo{author}{\bibfnamefont{R.~D.} \bibnamefont{Groot}} \bibnamefont{and}
  \bibinfo{author}{\bibfnamefont{P.~B.} \bibnamefont{Warren}},
  \bibinfo{journal}{J. Chem. Phys.} \textbf{\bibinfo{volume}{107}},
  \bibinfo{pages}{4423} (\bibinfo{year}{1997}).

\bibitem[{\citenamefont{Smiatek et~al.}(2009)\citenamefont{Smiatek, Sega, Holm,
  Schiller, and Schmid}}]{Smiatek2009}
\bibinfo{author}{\bibfnamefont{J.}~\bibnamefont{Smiatek}},
  \bibinfo{author}{\bibfnamefont{M.}~\bibnamefont{Sega}},
  \bibinfo{author}{\bibfnamefont{C.}~\bibnamefont{Holm}},
  \bibinfo{author}{\bibfnamefont{U.~D.} \bibnamefont{Schiller}},
  \bibnamefont{and} \bibinfo{author}{\bibfnamefont{F.}~\bibnamefont{Schmid}},
  \bibinfo{journal}{J. Chem. Phys.} \textbf{\bibinfo{volume}{130}},
  \bibinfo{eid}{244702} (\bibinfo{year}{2009}).

\bibitem[{\citenamefont{Zhou and Schmid}(2012)}]{2012_ac}
\bibinfo{author}{\bibfnamefont{J.}~\bibnamefont{Zhou}} \bibnamefont{and}
  \bibinfo{author}{\bibfnamefont{F.}~\bibnamefont{Schmid}},
  \bibinfo{journal}{J. Phys.: Condens. Matter} \textbf{\bibinfo{volume}{24}},
  \bibinfo{pages}{464112} (\bibinfo{year}{2012}).

\bibitem[{\citenamefont{Soddemann et~al.}(2003)\citenamefont{Soddemann,
  D\"unweg, and Kremer}}]{Soddemann2003}
\bibinfo{author}{\bibfnamefont{T.}~\bibnamefont{Soddemann}},
  \bibinfo{author}{\bibfnamefont{B.}~\bibnamefont{D\"unweg}}, \bibnamefont{and}
  \bibinfo{author}{\bibfnamefont{K.}~\bibnamefont{Kremer}},
  \bibinfo{journal}{Phys. Rev. E} \textbf{\bibinfo{volume}{68}},
  \bibinfo{pages}{046702} (\bibinfo{year}{2003}).

\bibitem[{\citenamefont{Weeks et~al.}(1971)\citenamefont{Weeks, Chandler, and
  Andersen}}]{WCA}
\bibinfo{author}{\bibfnamefont{J.~D.} \bibnamefont{Weeks}},
  \bibinfo{author}{\bibfnamefont{D.}~\bibnamefont{Chandler}}, \bibnamefont{and}
  \bibinfo{author}{\bibfnamefont{H.~C.} \bibnamefont{Andersen}},
  \bibinfo{journal}{J. Chem. Phys.} \textbf{\bibinfo{volume}{54}},
  \bibinfo{pages}{5237} (\bibinfo{year}{1971}).

\bibitem[{\citenamefont{Wright}(2007)}]{Wright}
\bibinfo{author}{\bibfnamefont{M.~R.} \bibnamefont{Wright}},
  \emph{\bibinfo{title}{An Introduction to Aqueous Electrolyte Solutions}}
  (\bibinfo{publisher}{Wiley}, \bibinfo{address}{Chichester},
  \bibinfo{year}{2007}).

\bibitem[{\citenamefont{Hasimoto}(1959)}]{Hasimoto1959}
\bibinfo{author}{\bibfnamefont{H.}~\bibnamefont{Hasimoto}},
  \bibinfo{journal}{J. Fluid Mech.} \textbf{\bibinfo{volume}{5}},
  \bibinfo{pages}{317} (\bibinfo{year}{1959}).

\bibitem[{\citenamefont{Limbach et~al.}(2006)\citenamefont{Limbach, Arnold,
  Mann, and Holm}}]{ESPResSo}
\bibinfo{author}{\bibfnamefont{H.}~\bibnamefont{Limbach}},
  \bibinfo{author}{\bibfnamefont{A.}~\bibnamefont{Arnold}},
  \bibinfo{author}{\bibfnamefont{B.}~\bibnamefont{Mann}}, \bibnamefont{and}
  \bibinfo{author}{\bibfnamefont{C.}~\bibnamefont{Holm}},
  \bibinfo{journal}{Comp. Phys. Comm.} \textbf{\bibinfo{volume}{174}},
  \bibinfo{pages}{704} (\bibinfo{year}{2006}).

\bibitem[{\citenamefont{Hockney and Eastwood}(1988)}]{HockneyEastwood}
\bibinfo{author}{\bibfnamefont{R.}~\bibnamefont{Hockney}} \bibnamefont{and}
  \bibinfo{author}{\bibfnamefont{J.}~\bibnamefont{Eastwood}},
  \emph{\bibinfo{title}{Computer Simulation Using Particles}}
  (\bibinfo{publisher}{Adam Hilger}, \bibinfo{address}{Bristol},
  \bibinfo{year}{1988}).

\bibitem[{\citenamefont{Deserno and Holm}(1998{\natexlab{a}})}]{Deserno1998}
\bibinfo{author}{\bibfnamefont{M.}~\bibnamefont{Deserno}} \bibnamefont{and}
  \bibinfo{author}{\bibfnamefont{C.}~\bibnamefont{Holm}}, \bibinfo{journal}{J.
  Chem. Phys.} \textbf{\bibinfo{volume}{109}}, \bibinfo{pages}{7678}
  (\bibinfo{year}{1998}{\natexlab{a}}).

\bibitem[{\citenamefont{Deserno and Holm}(1998{\natexlab{b}})}]{Deserno1998a}
\bibinfo{author}{\bibfnamefont{M.}~\bibnamefont{Deserno}} \bibnamefont{and}
  \bibinfo{author}{\bibfnamefont{C.}~\bibnamefont{Holm}}, \bibinfo{journal}{J.
  Chem. Phys.} \textbf{\bibinfo{volume}{109}}, \bibinfo{pages}{7694}
  (\bibinfo{year}{1998}{\natexlab{b}}).

\bibitem[{\citenamefont{Verlet}(1967)}]{Verlet1967}
\bibinfo{author}{\bibfnamefont{L.}~\bibnamefont{Verlet}},
  \bibinfo{journal}{Phys. Rev.} \textbf{\bibinfo{volume}{159}},
  \bibinfo{pages}{98} (\bibinfo{year}{1967}).

\bibitem[{\citenamefont{Swope et~al.}(1982)\citenamefont{Swope, Andersen,
  Berens, and Wilson}}]{Swope1982}
\bibinfo{author}{\bibfnamefont{W.~C.} \bibnamefont{Swope}},
  \bibinfo{author}{\bibfnamefont{H.~C.} \bibnamefont{Andersen}},
  \bibinfo{author}{\bibfnamefont{P.~H.} \bibnamefont{Berens}},
  \bibnamefont{and} \bibinfo{author}{\bibfnamefont{K.~R.}
  \bibnamefont{Wilson}}, \bibinfo{journal}{J. Chem. Phys.}
  \textbf{\bibinfo{volume}{76}}, \bibinfo{pages}{637} (\bibinfo{year}{1982}).

\bibitem[{\citenamefont{Frenkel and Smit}(2002)}]{FrenkelSmit}
\bibinfo{author}{\bibfnamefont{D.}~\bibnamefont{Frenkel}} \bibnamefont{and}
  \bibinfo{author}{\bibfnamefont{B.}~\bibnamefont{Smit}},
  \emph{\bibinfo{title}{Understanding Molecular Simulation}}
  (\bibinfo{publisher}{Academic Press}, \bibinfo{year}{2002}),
  \bibinfo{edition}{2nd} ed.

\bibitem[{\citenamefont{Padding and Louis}(2006)}]{Padding2006}
\bibinfo{author}{\bibfnamefont{J.~T.} \bibnamefont{Padding}} \bibnamefont{and}
  \bibinfo{author}{\bibfnamefont{A.~A.} \bibnamefont{Louis}},
  \bibinfo{journal}{Phys. Rev. E} \textbf{\bibinfo{volume}{74}},
  \bibinfo{pages}{031402} (\bibinfo{year}{2006}).

\bibitem[{\citenamefont{Ohshima}(2006)}]{Ohshima}
\bibinfo{author}{\bibfnamefont{H.}~\bibnamefont{Ohshima}},
  \emph{\bibinfo{title}{Theory of Colloid and Interfacial Electric Phenomena}}
  (\bibinfo{publisher}{Academic Press}, \bibinfo{address}{Amsterdam},
  \bibinfo{year}{2006}).

\bibitem[{\citenamefont{Hill et~al.}(2003{\natexlab{b}})\citenamefont{Hill,
  Saville, and Russel}}]{Hill2003a}
\bibinfo{author}{\bibfnamefont{R.~J.} \bibnamefont{Hill}},
  \bibinfo{author}{\bibfnamefont{D.~A.} \bibnamefont{Saville}},
  \bibnamefont{and} \bibinfo{author}{\bibfnamefont{W.~B.}
  \bibnamefont{Russel}}, \bibinfo{journal}{J. Colloid Interface Sci.}
  \textbf{\bibinfo{volume}{258}}, \bibinfo{pages}{56}
  (\bibinfo{year}{2003}{\natexlab{b}}).

\bibitem[{\citenamefont{Loeb et~al.}(1961)\citenamefont{Loeb, Overbeek, and
  Wiersema}}]{LOW_table}
\bibinfo{author}{\bibfnamefont{A.~L.} \bibnamefont{Loeb}},
  \bibinfo{author}{\bibfnamefont{J.~T.~G.} \bibnamefont{Overbeek}},
  \bibnamefont{and} \bibinfo{author}{\bibfnamefont{P.~H.}
  \bibnamefont{Wiersema}}, \emph{\bibinfo{title}{The Electrical Double Layer
  around a Spherical Colloid Particle}} (\bibinfo{publisher}{MIT Press},
  \bibinfo{address}{Massachusetts}, \bibinfo{year}{1961}).

\bibitem[{\citenamefont{Ohshima et~al.}(1982)\citenamefont{Ohshima, Healy, and
  White}}]{Ohshima1982}
\bibinfo{author}{\bibfnamefont{H.}~\bibnamefont{Ohshima}},
  \bibinfo{author}{\bibfnamefont{T.}~\bibnamefont{Healy}}, \bibnamefont{and}
  \bibinfo{author}{\bibfnamefont{L.}~\bibnamefont{White}}, \bibinfo{journal}{J.
  Colloid Interface Sci.} \textbf{\bibinfo{volume}{90}}, \bibinfo{pages}{17}
  (\bibinfo{year}{1982}).

\bibitem[{\citenamefont{Baptista et~al.}(2009)\citenamefont{Baptista, Schmitz,
  and D\"unweg}}]{Baptista2009}
\bibinfo{author}{\bibfnamefont{M.}~\bibnamefont{Baptista}},
  \bibinfo{author}{\bibfnamefont{R.}~\bibnamefont{Schmitz}}, \bibnamefont{and}
  \bibinfo{author}{\bibfnamefont{B.}~\bibnamefont{D\"unweg}},
  \bibinfo{journal}{Phys. Rev. E} \textbf{\bibinfo{volume}{80}},
  \bibinfo{pages}{016705} (\bibinfo{year}{2009}).

\bibitem[{\citenamefont{Schmitz}(2011)}]{phd_Schmitz}
\bibinfo{author}{\bibfnamefont{R.}~\bibnamefont{Schmitz}}, Ph.D. thesis,
  \bibinfo{school}{Johannes Gutenberg-Universit\"at Mainz}
  (\bibinfo{year}{2011}).

\bibitem[{\citenamefont{Bikerman}(1940)}]{Bikerman1940}
\bibinfo{author}{\bibfnamefont{J.}~\bibnamefont{Bikerman}},
  \bibinfo{journal}{Trans. Faraday Soc.} \textbf{\bibinfo{volume}{35}},
  \bibinfo{pages}{154} (\bibinfo{year}{1940}).

\bibitem[{\citenamefont{O'Brien}(1986)}]{OBrien1986}
\bibinfo{author}{\bibfnamefont{R.~W.} \bibnamefont{O'Brien}},
  \bibinfo{journal}{J. Colloid Interface Sci.} \textbf{\bibinfo{volume}{113}},
  \bibinfo{pages}{81} (\bibinfo{year}{1986}).

\end{thebibliography}


\end{document}